**Bayesian optimization with experimental failure for high-throughput materials growth**

Yuki K. Wakabayashi,[1,*,†] Takuma Otsuka,[2,*,‡] Yoshiharu Krockenberger,[1] Hiroshi Sawada,[2] Yoshitaka Taniyasu,[1] and Hideki Yamamoto[1]

[1]*NTT Basic Research Laboratories, NTT Corporation, Atsugi, Kanagawa 243-0198, Japan*
[2]*NTT Communication Science Laboratories, NTT Corporation, Soraku-gun, Kyoto 619-0237, Japan*

[*]These authors contributed equally to this work.
[†]Corresponding author: yuuki.wakabayashi.we@hco.ntt.co.jp
[‡]Corresponding author: takuma.otsuka.uf@hco.ntt.co.jp

**Abstract**
A crucial problem in achieving innovative high-throughput materials growth with machine learning and automation techniques, such as Bayesian optimization (BO) and robotic experimentation, has been a lack of an appropriate way to handle missing data due to experimental failures. Here, we propose a new BO algorithm that complements the missing data in the optimization of materials growth parameters. The proposed method provides a flexible optimization algorithm capable of searching a wide multi-dimensional parameter space. We demonstrate the effectiveness of the method with simulated data as well as in its implementation for actual materials growth, namely machine-learning-assisted molecular beam epitaxy (ML-MBE) of $SrRuO_3$, which is widely used as a metallic electrode in oxide electronics. Through the exploitation and exploration in a wide three-dimensional parameter space, while complementing the missing data, we prepared tensile-strained $SrRuO_3$ film with a high residual resistivity ratio of 80.1, the highest among tensile-strained $SrRuO_3$ films ever reported, in only 35 MBE growth runs.

# INTRODUCTION

Recent advances in materials informatics exploiting machine learning techniques, such as Bayesian optimization (BO) and artificial neural networks, offer an excellent opportunity to accelerate materials research[1,2,3,4,5]. In particular, data-driven decision-making approaches have attained high-throughput in experiments where machine learning models are incrementally updated by newly measured data[6,7,8,9,10]. BO is a sample-efficient approach for global optimization[11]. It has proven itself useful to streamline the optimization of the materials synthesis conditions for bulk[12] and thin film materials[13]. Taking thin film growth as an example, the BO algorithm provides the growth conditions that should be examined in the next growth run, which enables automatic optimization of the growth parameters. In combination with automated growth[14,15,16] and characterization apparatuses, the entire growth process has been automated. These approaches are collectively referred to as autonomous materials synthesis[17,18,19,20].

The missing data problem is a general issue encountered in materials informatics when analyzing real materials data[21,22]. Since missing data are common in various materials databases, appropriate handling of the missing data is vital for accelerating materials research. Furthermore, this problem is also critical in optimizing the conditions for materials growth since it is caused in the growth parameter space when a target material cannot be obtained due to growth parameters that are far from optimal. One possible solution is to restrict the search space of growth parameters so that it does not include missing data. Such restriction may be empirically carried out based on the experience and intuition of researchers and/or the information available in databases and the literature. However, there is no guarantee that the optimal growth parameters for the target material exist in such a small parameter space. Therefore, to maximize the benefit of high-throughput materials growth, including autonomous materials synthesis, it is essential to search in a wide parameter space while complementing the missing data generated due to unsuccessful growth runs, e.g., when the designated phase is not formed.

This study proposes a new BO method capable of handling missing data even when a target material has not formed due to growth parameters that are far from optimal. We equalize the evaluation value for the missing data to the worst evaluation value available at that time. This imputation of experimental failure enables us to search a wide parameter space and avoid unstable parameter regions. We demonstrate the effectiveness of the BO method with experimental failure first by using virtual data for simulation, and subsequently through implementation for real materials growth, namely machine-learning-assisted molecular beam epitaxy (ML-MBE) of itinerant ferromagnetic perovskite $SrRuO_3$ thin films, where we used the residual resistivity ratio (RRR) as the evaluation value. We achieved the RRR of 80.1, the highest ever reported among tensile-strained $SrRuO_3$ films, through the exploitation and exploration in a wide three-dimensional parameter space in only 35 MBE growth runs—the tensile-strained $SrRuO_3$ thin films achieved by epitaxial strain showed higher Curie temperature than those of bulk or compressive-strained films[23]. The proposed method provides a flexible optimization algorithm for a wide multi-dimensional parameter space that assumes experimental failure, and it will enhance the efficiency of high-throughput materials growth and autonomous materials growth.

**RESULTS AND DISCUSSION**
**Simulation with virtual data**
Our parameter search problem is described as follows. Given a multidimensional parameter denoted by vector $\mathbf{x}$, an experimental trial returns its evaluation $y$. In materials growth optimization, $\mathbf{x}$ and $y$ represent growth parameters and physical property that evaluate grown materials, respectively. Examples of physical properties include electrical resistance, X-ray diffraction intensity, and so on. The choice of evaluation metrics should be determined by the purpose of the study. We assume an additive observation noise $y = S(\mathbf{x}) + e$, where $e$ follows the normal distribution as $e \sim \mathcal{N}(0, \sigma^2)$. This noise in $y$ corresponds to experimental fluctuations of the physical property currently under focus in materials prepared nominally under the same conditions. Here, the relationship between the parameter and evaluation, denoted by function $S(\cdot)$, is unknown a-priori. Our goal is to find $\mathbf{x}^*$ such that its corresponding evaluation $S(\mathbf{x}^*)$ is maximized. Since the underlying $S$ is unknown, we sequentially carry out experimental trials with a parameter that is likely to return a better evaluation given past observations. That is, we choose a parameter $\mathbf{x}_n$ such that its result $y_n \approx S(\mathbf{x}_n)$ is predicted to be high given the data observed so far $\{(\mathbf{x}_1, y_1), \ldots, (\mathbf{x}_{n-1}, y_{n-1})\}$, where $\mathbf{x}_n$ and $y_n$ are $n$th parameter and $n$th evaluation values, respectively. Bayesian optimization is adopted for this sequential parameter search (see Methods section "Bayesian optimization with experimental failure" for details).

Our technical challenge is how to handle experimental failures. Specifically, evaluation $y_n$ may not always be returned for a specified parameter $\mathbf{x}_n$. To meet this challenge, we need to satisfy two requirements. First, the optimization procedure should avoid subsequent failures if the target material has not been formed under certain synthesis conditions. Second, even if a certain parameter $\mathbf{x}_n$ turn out to be a failure, the prediction model should be updated. The lack of evaluation $y_n$ provides some information. For example, if $\mathbf{x}_n$ led to a failure, parameters in a distance from the failed $\mathbf{x}_n$ should be favored because $\mathbf{x}_n$ is far from optimal parameter region or $\mathbf{x}_n$ is a condition in which another undesired material is stabilized. Thus, a failure can guide the sequential optimization by encouraging the exploration of other parametric regions.

This paper seeks two approaches to cope with experimental failures. The first approach is called the *floor padding* trick: when the experiment turns out a failure given parameter $\mathbf{x}_n$, the floor padding trick uses the worst value observed so far, namely the $y_n$ value is complemented by $\min_{1 \leq i < n} y_i$. This simple trick provides the search algorithm with information that the attempted parameter $\mathbf{x}_n$ worked negatively. At the same time, this method is adaptive and automatic. The past experiments determine how bad a failure should count; in contrast, a naïve alternative is to give a predetermined constant value to failures. This may require careful tuning of the padding constant. The floor padding trick fulfills both requirements discussed above: the worst evaluation helps avoid parameters near the failure and updates the prediction model as well. The other approach is the *binary classifier* of failures, which, in addition to a prediction model for the evaluation value $y_n$, we employ to predict whether a given parameter $\mathbf{x}_n$ leads to a failure or not. The binary classifier meets the first requirement to avoid subsequent failures. Nevertheless, the second requirement is yet to be addressed since the binary classifier may not affect the evaluation prediction when we employ distinct models. Thus, the floor padding or naïve constant padding is combined with the binary classifier to update the prediction model accordingly.

We designed several methods by incorporating the floor padding trick and the binary classifier. These methods are compared using a simulated function, where simulation means the evaluation is calculated by artificial functions (see Methods section "Simulated functions" for details). They are designed to investigate the efficiency of the various combination of the floor padding trick (abbreviated as 'F') and the binary classifier (abbreviated as 'B'), as summarized in Table I. When F is active, the method uses the floor padding for experimental failures, while the baseline uses a predetermined constant value without F. Similarly, when B is active, the method constructs a binary classifier[24] to predict failures in addition to the prediction model of evaluation, both of which are based on the Gaussian process[25] (see Methods section "Bayesian optimization with experimental failure" for details). In this work, each method searched the two types of simulated parameter spaces (Fig. 1) until 100 observations were queried. This search process was repeated five times to account for randomness. For each process, five search points were randomly chosen as the initial observations. The observation noise was set to $\sigma^2 = 0.005$.

Figure 2 shows the parameter search results for the Circle function. Each curve indicates the best evaluation value averaged over five runs as a function of the number of observations, whereas the shaded areas represent the best and worst evaluation among the five runs for each method. A curve that rises with fewer observations indicates a better search algorithm since it requires fewer resources before reaching a high evaluation. A vertically narrow shaded area means that the method performs robustly against randomness in the search process, such as the initial choice of parameters and observation noise. First, we observe that the choice of a constant value to replace failed evaluations affects the curves in Fig. 2b. In particular, the slope of the curves at the early stage is sensitive to the choice of the constat; for example, baseline @0 shows a quick improvement at first while @−1 gives slower improvements. Appropriate tuning of this value may be difficult in general and dependent on the experience and knowledge of individual researchers. As shown in Fig. 2a, the floor padding trick of method F demonstrates an initial improvement as quick as @0 without careful tuning of the constant. With that said, the final average evaluation of F is suboptimal compared to the best @−1. Second, method FB with the floor padding and binary classifier combined results in slower improvements of evaluation (Fig. 2a). The binary classifier alleviates the sensitivity to the choice of the constant value for handling failures in that the discrepancy between the curves in Fig. 2c is suppressed compared with that in Fig. 2b. However, the initial improvements in the evaluation of B@0 and B@−1 are inferior to that of F (Fig. 2a), and the final evaluation is also exceeded by @−1 (Fig. 2b).

Figure 3 shows the results for the Hole function. The overall tendency of the two tricks is similar to that of Circle: the floor padding leads to quick improvements of the evaluation in the early part of the search process (Fig. 3a). We observe sensitivity to the choice of padding constant in Fig. 3b: the baseline of @−1 quickly improved the evaluation and reached a high evaluation on average at the end, whereas @0 struggled to find good parameters. Note that a favorable choice of the padding constant depends on the function: baseline @0 showed quicker improvements for the Circle function (Fig. 2b), while @−1 performed better for the Hole function (Fig. 3b). Method F with the floor padding trick gave competitive curves at early stages for both functions compared to the baseline methods. Another similarity to the Circle result is that method B shows less sensitivity to the choice of constant in Fig. 3c and , at the same time, slower improvement compared to method F in Fig. 3a or @−1 in Fig. 3b with a well-chosen constant. Method FB, the combination of the floor padding and binary classifier, was comparable to method F, as shown in Fig. 3a. Considering the evaluation of the highest peak is approximately 1.85 while the rest of the extrema are 1.52 (see Methods section "Simulated functions" for details), F and FB often reached the optimum and at least found one of the extrema at the end in the five runs.

Our simulation experiments confirmed the efficacy of the floor padding: this trick is beneficial for initial quick search progress and sidesteps the difficult choice of padding constants. As for the binary classifier, this approach also alleviates the sensitivity to the choice of constant, though the search progress can be slowed. Furthermore, combining it with the floor padding was not as effective as expected. Thus, we adopted method F for the ML-MBE growth of $SrRuO_3$ films to make the most of limited resources for experiments.

**Application for ML-MBE of $SrRuO_3$ films**

To demonstrate the applicability and effectiveness of the BO method with experimental failure to actual growth of materials, we applied method F to our recently developed ML-MBE[13] of $SrRuO_3$ films on $DyScO_3$ (110) substrates. The itinerant 4$d$ ferromagnetic perovskite $SrRuO_3$ is one of the most promising materials for oxide electronics owing to its high metallic conduction, chemical stability, compatibility with other perovskite oxides, and ferromagnetism with strong uniaxial magnetic anisotropy[26,27,28,29,30,31,32,33,34,35]. In addition, the recent observation of the Weyl fermions in $SrRuO_3$ points to this material as an appropriate platform to integrate two emerging fields: topology in condensed matter and oxide electronics[36,37,38,39,40]. The RRR value, defined as the ratio of resistivity at 300 K [$\rho(300\ K)$] to that at 4 K [$\rho(4\ K)$], is a good measure of the purity of metallic systems, and accordingly, the quality of single-crystalline $SrRuO_3$ thin films[31,35,41,42,43]. For practical applications of $SrRuO_3$, such as electrodes in dynamic random access memory application, high-quality $SrRuO_3$ films are necessary[44,45]. In terms of physical interest, high-quality $SrRuO_3$ films are also indispensable because only $SrRuO_3$ films with high RRR values above 20 have enabled the observation of quantum transport of Weyl fermions[37,38]. Thus, we adopted RRR as the evaluation value.

Figure 4 shows the flow of the ML-MBE growth using the BO with experimental failure (see Methods section "ML-MBE growth and sample characterizations" for details). For the growth of high-quality $SrRuO_3$, fine tuning of the growth conditions (the supplied Ru/Sr flux ratio, growth temperature, and $O_3$-nozzle-to-substrate distance) is important[13,45]. In a previous study[13], to simplify the intricate search space of entangled growth conditions, we ran the BO for a single growth condition while keeping the other conditions fixed. In addition, the search range was reduced to the growth parameter range within which the $SrRuO_3$ phase had formed. In contrast, in the present study, we ran the BO algorithm in the three-dimensional space directly. Following the F method, when the SrRuO$_3$ phase had not formed, we defined the RRR value of those samples to be the worst experimental RRR value by that time, instead of reducing the search range. The search ranges for the Ru flux rate, growth temperature, and O$_3$-nozzle-to-substrate distance were 0.25–0.50 Å/s, 700–900°C, and 10–50 mm, respectively. Here, the Ru/Sr ratio is determined by the Ru flux ratio as the Sr flux rate was fixed at 0.98 Å/s. The O$_3$-nozzle-to-substrate distance is a parameter for oxidation strength.

Figure 5 shows how the BO algorithm predicts RRR values with unseen parameter configurations and acquires new data points. The process starts with five random initial growth parameters [Fig. 5a] and gains experimental RRR values for the updated GPR model with 18 [Fig. 5b] and 37 [Fig. 5c] samples. Two-dimensional plots of the predicted RRR, $s$, and EI values at the O$_3$-nozzle-to-substrate distance, at which the highest EI value was obtained, are shown in the lower panels [Figs. 5d–5l]. Among the five random initial growth parameters, the SrRuO$_3$ phase had not formed at the Ru flux rate = 0.47 Å/s, growth temperature = 832°C, and the O$_3$-nozzle-to-substrate distance = 25 mm (Fig. 5a). Thus, we defined the RRR value of this sample to be the worst experimental one at that time (13.1). This imputation of experimental failure enabled direct search of the wide three-dimensional parameter space. The prediction results from the five initial samples yielded the highest EI at a O$_3$ nozzle-substrate distance 1.5 mm larger than that for the highest experimental RRR value at that time (Fig. 5a). The RRR obtained at the next set of growth parameters (RRR = 35) was slightly larger than the highest RRR value of the five initial samples (RRR = 33.4). The region with relatively small $s$ of the predicted RRR became larger as the number of experimental samples increased from five to 37 [Figs. 5g-5i], indicating that the accuracy of the prediction had increased. The small $s$ values result in lower EI values [Figs. 5j-5l]. This suggests that we have only a limited chance to improve the RRR value by further modifying the growth parameters. Through this exploitation and exploration of the materials growth data with experimental failure in the three-dimensional parameter space, the highest RRR value increased and reached over 80 in only 35 MBE growth runs (Fig. 6). The highest experimental RRR of 80.1, was achieved at the Ru flux = 0.365 Å/s, growth temperature = 826 °C, and O$_3$-nozzle-to-substrate distance = 22 mm. This is the highest RRR ever reported among tensile-strained SrRuO$_3$ films[23,31]. The achievement of the target material with the highest conductivity in such a small number of optimizations demonstrates the effectiveness of this method utilizing the floor padding trick for high-throughput materials growth.

In summary, we presented a new BO algorithm which complements the missing data. We designed several methods by combining the floor padding trick and the binary classifier, and compared them by simulation experiments. We found the efficacy of the floor padding trick, while the binary classifier decelerate the search in the early stages. The imputation of missing data to the worst experimental RRR value at that time allows us to search a wide parameter space directly and will enhance the time and cost efficiency of materials growth and autonomous materials synthesis. With regard to the growth parameter optimization for the growth of high-quality tensile-strained SrRuO$_3$ thin films, the experimental RRR of 80.1 was achieved in only 35 ML-MBE runs with the floor padding trick. The proposed method, providing a flexible optimization algorithm for a wide multi-dimensional parameter space that assumes experimental failure, will play an essential role in the growth of various materials.

**METHODS**
**Bayesian optimization with experimental failure**
Here, we describe our parameter search method based on Bayesian optimization. Our notation uses italic characters to denote scalar values, bold lower-case symbols for vectors, and bold upper-case characters for matrices. Supplementary Fig. 1 outlines our method. Let $x_{n,d}$ denote the *d*th element of the parameter in the *n*th trial. That is, the *D*-dimensional parameter is written as a column vector $\mathbf{x}_n = [x_{n,1}, \ldots, x_{n,D}]^\top$, where $\cdot^\top$ is the transpose operator. We assume the search space is bounded; namely, every element has its minimum and maximum values, $\underline{x}_d$ and $\overline{x}_d$, respectively. The value of *n*th evaluation $y_n$ is a real number when the experiment is successful, but is denoted as $\phi$ otherwise. In our implementation, $\phi =$ NaN (not a number) for experimental failure. The parameter is first normalized as

$$\tilde{x}_{n,d} = \frac{x_{n,d} - \underline{x}_d}{\overline{x}_d - \underline{x}_d}, \tag{1}$$

such that all elements in $\tilde{\mathbf{x}}_n$ fall between 0 and 1. This facilitates the training of the Gaussian process used as the prediction model. The floor padding trick preprocesses the evaluations as follows. Let $y_1, \ldots, y_{n-1}$ be the observed evaluations so far. The failed evaluations are replaced with the worst successful value as:

$$\tilde{y}_{n'} = \begin{cases} y_{n'} & \text{if } y_{n'} \neq \phi, \\ \min_{1 \leq i < n} \tilde{y}_i & \text{if } y_{n'} = \phi, \end{cases} \tag{2}$$

for $n' = 1, \ldots, n-1$. If $y_1 = \phi$, then $\tilde{y}_1 = 0$ or any other reasonable guess of low evaluation. Supplementary Fig. 2 illustrates the effect of the floor padding trick compared with the methods with constant values for failed evaluations. The predicted function surface around $[0.6, 0.2]^\top$ is affected by the choice of the complementary value — the predicted value is significantly penalized when the failure is treated by low value $\tilde{y} = -1$ (Supplementary Fig. 2c), whereas the floor padding trick (Supplementary Fig. 2a) gives a milder decrease in the prediction around failures indicated by '−' marks.

We start the search by trying a random initial point $\mathbf{x}_1$ to obtain the first evaluation $y_1$. We may optionally start with multiple initial points depending on the budget for the total number of experiments. Then, we iterate the update of the prediction model, decide a parameter for the next experiment, and then receive its evaluation through the actual experimental process.

Given the preprocessed $n-1$ observations $\mathcal{D}_{n-1} \coloneqq \{\tilde{\mathbf{x}}_i, \tilde{y}_i\}_{i=1}^{n-1}$ so far, we fit the Gaussian process to predict the evaluation of the unexplored parameter region and to decide which parameter to try in the next step. The predictive evaluation $\tilde{y}'$ at normalized parameter $\tilde{\mathbf{x}}'$ given $\mathcal{D}_{n-1}$ is specified as the following normal distribution[46]: $p(\tilde{y}'|\tilde{\mathbf{x}}', \mathcal{D}_{n-1}) = \mathcal{N}(m(\tilde{\mathbf{x}}'), s^2(\tilde{\mathbf{x}}'))$, where the mean and variance are calculated as follows.

$$m(\tilde{\mathbf{x}}') = \mathbf{k}_{n-1}(\tilde{\mathbf{x}}')^\top (\sigma^2 \mathbf{I} + \mathbf{K}_{n-1})^{-1} \tilde{\mathbf{y}}_{n-1}, \tag{3}$$
$$s^2(\tilde{\mathbf{x}}') = \sigma^2 + k(\tilde{\mathbf{x}}', \tilde{\mathbf{x}}') - \mathbf{k}_{n-1}(\tilde{\mathbf{x}}')^\top (\sigma^2 \mathbf{I} + \mathbf{K}_{n-1})^{-1} \mathbf{k}_{n-1}(\tilde{\mathbf{x}}'). \tag{4}$$

Here, $k(\tilde{\mathbf{x}}', \tilde{\mathbf{x}}')$ is the kernel function, $\mathbf{k}_{n-1}(\tilde{\mathbf{x}}')$ is the kernel vector, $\mathbf{K}_{n-1}$ is the kernel matrix, $\tilde{\mathbf{y}}_{n-1} = [\tilde{y}_1, \ldots, \tilde{y}_{n-1}]^\top$, and $\mathbf{I}$ is an identity matrix of appropriate size. The mean $m(\tilde{\mathbf{x}}')$ and variance $s(\tilde{\mathbf{x}}')$ of the predictive distribution of Eqs. (3) and (4) are derived through matrix and vector operations. The kernel vector is defined with the kernel function $k(\cdot,\cdot)$: $\mathbf{k}_{n-1}(\tilde{\mathbf{x}}') = [k(\tilde{\mathbf{x}}', \tilde{\mathbf{x}}_1), \ldots, k(\tilde{\mathbf{x}}', \tilde{\mathbf{x}}_{n-1})]^\top$. The $i$th row and $j$th column of the matrix $\mathbf{K}_{n-1}$ is defined as $k_{ij} = k(\tilde{\mathbf{x}}_i, \tilde{\mathbf{x}}_j)$. The kernel represents the similarity between the two parameters. When $\tilde{\mathbf{x}}_i$ and $\tilde{\mathbf{x}}_j$ are close to each other, $k_{ij}$ amounts to a large value, which leads to a high correlation between $\tilde{y}_i$ and $\tilde{y}_j$. For Bayesian optimization, the radial basis function (RBF) kernel and Matérn $5/2$ kernel are popular choices. When the distance is given by $\Delta_{ij} = \sqrt{\sum_{d=1}^{D} \frac{(\tilde{x}_{i,d} - \tilde{x}_{j,d})^2}{s_d}}$, the RBF kernel is defined as $k_{ij} = A\exp(-\Delta_{ij}^2)$, and Matérn $5/2$ kernel is $k_{ij} = A\left(1 + \sqrt{5}\Delta_{ij} + \frac{5}{3}\Delta_{ij}^2\right)\exp(-\sqrt{5}\Delta_{ij})$. Here, $A > 0$ and $s_d > 0$ are hyperparameters of the kernel and usually tuned to maximize the likelihood of observation $\mathcal{D}_{n-1}$[25].

Once the prediction model is constructed, we will choose a parameter that maximizes the possibility to produce a high evaluation value in the next experiment. We use the expected improvement (EI) criterion. EI is defined as follows:

$$a_{\text{EI}}(\tilde{\mathbf{x}}') = \mathbb{E}_{p(\tilde{y}'|\tilde{\mathbf{x}}', \mathcal{D}_{n-1})}\left[(\tilde{y}' - \overline{y}_{n-1})\mathbb{I}_{\tilde{y}' \geq \overline{y}_{n-1}}\right], \quad (5)$$

where $\mathbb{I}_{\tilde{y}' \geq \overline{y}_{n-1}}$ is the indicator function, which equals 1 when $\tilde{y}' \geq \overline{y}_{n-1}$ and 0 otherwise, and $\overline{y}_{n-1} = \max_{1 \leq i < n} \tilde{y}_n$ is the best evaluation so far. The intuition of EI is that it measures the expected gain over the best-observed evaluation. Thus, our next parameter is the maximizer of $a_{\text{EI}}(\tilde{\mathbf{x}}')$ where the search space is $0 \leq \tilde{x}'_d \leq 1$ for all $d$.

The EI function is modified when the binary classifier is adopted to predict the experimental failures. The modified criterion takes the form

$$a_{\text{EI,B}}(\tilde{\mathbf{x}}') = a_{\text{EI}}(\tilde{\mathbf{x}}')p(y' \neq \phi|\tilde{\mathbf{x}}', \mathcal{D}_{n-1}). \quad (6)$$

This is the product of the original EI and the predicted probability of $\tilde{\mathbf{x}}'$ not resulting in a failure $\phi$. By assuming the occurrence of experimental failure is independent from the evaluation value, this modified EI amounts to the expectation of the experiment being successful with its gain in the evaluation from the best one obtained so far. The assumption is justified by obtaining a distinct classifier and evaluation prediction model.

This formulation is an extension of the constrained BO[24], where the objective is given as

$$\max_{\mathbf{x}} S(\mathbf{x}) \text{ s.t. } C(\mathbf{x}) \leq c, \quad (7)$$

where $C(\mathbf{x}) \leq c$ is a constraint. The constraint function $C(\mathbf{x})$ is unknown but observable: an observation consists of $(\mathbf{x}, y, z)$ with $y = S(\mathbf{x})$ and $z = C(\mathbf{x})$ (the observation noise was omitted for clarity). Two Gaussian processes are considered to predict $y'$ and $z'$ for unseen $\mathbf{x}'$. The resulting EI of the constrained BO[24] is the product of expected improvement in $y'$ and the probability of $z' \leq c$.

The binary classifier is the replacement of the constraint part. Instead of handling the continuous variable $z$, the classifier fits with the binary outcome: $y = \phi$ or $y \neq \phi$. We can still use the Gaussian process with the same kernel explained above, while variational inference is employed for the learning to approximate the likelihood of binary observations[25]. Supplementary Fig. 3 shows how the EI function is modified. The probability of success $p(y' \neq \phi | \tilde{\mathbf{x}}', \mathcal{D}_{n-1})$ is lowered by failed observations (Supplementary Fig. 3a). This directly discourages the exploration of failed regions in subsequent trials. Because of the approximate variational inference, the probability estimated by the binary classifier is not *sharp*; that is, it is not as high as 1 around successful observations and not as low as 0 around failures. This attenuates the variation in the EI function, as shown in Supplementary Figs. 3b and 3c.

**Simulated functions**
This section explains the setup of functions used in the simulation experiments. Figure 1 depicts two functions. The search space is two-dimensional between -1 and 1 for both functions. The Circle function fails when $x_1^2 + x_2^2 > 1$, whereas the Hole function fails when $|x_1| < L$ and $|x_2| < L$, or $x_1^2 + x_2^2 > 1$. With a slight abuse of notation, subscript $x_d$ means the *d*th element of the parameter by omitting the index of observation. Hole length is set as $L = \frac{\sqrt{\pi - 2}}{2}$ such that the area of failure is half of the search space. The Circle function is defined to have four peaks:

$$S_{\text{Circle}}(\mathbf{x}) = \frac{3}{2}\exp(-G_0(\mathbf{x})) + \sum_{i=1}^{3}\exp(-G_i(\mathbf{x})), \tag{8}$$

where $G_i(\mathbf{x}) = \sum_{d=1}^{2} a_{id} |x_d - c_{id}|$ with center
$$\mathbf{c}_0 = [0.7, 0]^\top, \mathbf{c}_1 = [0, 0.7]^\top, \mathbf{c}_2 = [-0.7, 0]^\top, \mathbf{c}_3 = [0, -0.7]^\top, \tag{9}$$
and scale being
$$\mathbf{a}_0 = [5,1]^\top, \mathbf{a}_1 = [1,5]^\top, \mathbf{a}_2 = [5,1]^\top, \mathbf{a}_3 = [1,5]^\top. \tag{10}$$

The peaks are connected with ridges, which may facilitate finding the top by tracing the ridges. The maximum value is 1.52 at $\mathbf{x} = [0.7, 0]^\top$. The Hole function is similarly defined but the ridges are tilted:

$$S_{\text{Hole}}(\mathbf{x}) = \frac{3}{2}\exp(-H_0(\mathbf{x})) + \sum_{i=1}^{3}\exp(-H_i(\mathbf{x})), \tag{12}$$

where $H_i(\mathbf{x}) = \sum_{d=1}^{2} a_{id} |z_{id}|$ with rotated vector
$$\mathbf{z}_i = \mathbf{R}(\mathbf{x} - \mathbf{c}_i'), \text{with } \mathbf{R} = \frac{1}{\sqrt{2}}\begin{bmatrix} 1 & -1 \\ 1 & 1 \end{bmatrix}. \tag{13}$$

Center coordinates were slightly moved to avoid touching the failure boundary:
$$\mathbf{c}_0' = [0.75, 0]^\top, \mathbf{c}_1' = [0, 0.75]^\top, \mathbf{c}_2' = [-0.75, 0]^\top, \mathbf{c}_3' = [0, -0.75]^\top,$$
whereas scale $\mathbf{a}_i$ was the same as that used for the Circle function. Since the peaks are disconnected, the parameter search may add some difficulty. The maximum value is 1.85 at $\mathbf{x} = [0.75, 0]^\top$.

**ML-MBE growth and sample characterizations**
Epitaxial tensile-strained SRO films with a thickness of 60 nm were grown on $DyScO_3$ (DSO) (110) substrates in a custom-designed molecular beam epitaxy (MBE) system with multiple e-beam evaporators (Fig. 4a). Rare-earth scandate $DyScO_3$ has a $GdFeO_3$ structure, a distorted perovskite structure with the (110) face corresponding to the pseudocubic (001) face[23]. Since the in-plane lattice parameter for the pseudocubic (001) face of $DyScO_3$ (3.944 Å) is larger than that of bulk $SrRuO_3$ (3.93 Å), the $SrRuO_3$ films epitaxially grown on the $DyScO_3$ substrate is tensile strained. Detailed information about the MBE system is described elsewhere[47–49]. We precisely controlled the elemental fluxes even for elements with high melting points, *e.g.*, Ru (2,250°C), by monitoring the flux rates with an electron-impact-emission-spectroscopy sensor and feeding the results back to the power supplies for the e-beam evaporators. The oxidation during growth was carried out with ozone ($O_3$) gas (~15% $O_3$ + 85% $O_2$) introduced through an alumina nozzle pointed at the substrate. For the high-quality $SrRuO_3$ growth, fine tuning of the growth conditions (the ratio of the Ru flux to the Sr flux, growth temperature, and local ozone pressure at the growth surface) is important[13,45]. To systematically change the Ru flux ratio to the Sr flux, we changed the Ru flux while keeping the Sr flux at 0.98 Å/s. The growth temperature was controlled by the heater shown in Fig. 4a. We can adjust the local ozone pressure at the growth surface by changing the $O_3$-nozzle-to-substrate distance while keeping the flow rate of $O_3$ gas at ~2 sccm. We ran the BO algorithm in the three-dimensional space. The search ranges for the Ru flux rate, growth temperature, and $O_3$-nozzle-to-substrate distance were 0.25–0.50 Å/s, from 700–900°C, and 10–50 mm, respectively. We searched equally spaced grid points for each parameter. The number and corresponding intervals of the respective quantities were 50 (0.005 Å/s interval), 100 (2°C interval), and 80 (0.5 mm interval). The experimental errors of the actual values from the nominal values for the respective growth parameters are typically ±0.005 Å/s, ±1°C, and ±0.2 mm, respectively. Since the three-dimensional parameter space consisted of 400,000 (50×100×80) points, performing a trial for the entire space in a point-by-point manner was unrealistic, as only several runs can be carried out per day with a typical MBE system. The crystal structure of the films was monitored by *in-situ* reflection high-energy electron diffraction (RHEED) after the growth. When diffractions from the $SrRuO_3$ phase were indiscernible and/or diffractions from SrO or $RuO_2$ precipitates (impurity phases) appeared, we defined the RRR value of those samples to be the worst experimental RRR value by that time. To determine the RRR of the samples, their electrical resistivity was measured using a standard four-probe method with Ag electrodes deposited on the $SrRuO_3$ surface without any additional processing. The distance between the two voltage electrodes was 2 mm.

Here, a black-box function RRR = S(**x**) is the target function specific to our $SrRuO_3$ films, and **x** represents the growth parameters (Ru flux rate, growth temperature, and $O_3$-nozzle-to-substrate distance). We used a data set $\mathcal{D}_{n-1} := \{\tilde{\mathbf{x}}_i, \tilde{y}_i\}_{i=1}^{n-1}$ (Fig. 4c) obtained from past $n-1$ MBE growths and RRR measurements (Fig. 4b) of $SrRuO_3$ films to construct a model to predict the value of S(**x**) at an unseen **x**. To this end, we use Gaussian process regression (GPR) to estimate the mean *m* and variance *s* at an arbitrary parameter value **x** (see Methods section "Bayesian optimization with experimental failure" for details). Specifically, GPR predicts the value of S(**x**) as a Gaussian-distributed variable $\mathcal{N}(m(\mathbf{x}), s^2(\mathbf{x}))$, where *m* and *s* depend on **x** and $n-1$ observations. In short, $m(\mathbf{x})$ represents the expected value of RRR and $s(\mathbf{x})$ represents the uncertainty of RRR at **x**. To consider the inherent noise in the RRR of $SrRuO_3$ films grown under nominally the same conditions, the variance of the observation noise $\sigma^2$ of the GPR model was set to 0.02 or 0.002. In our implementation, we used the same kernel function (Matérn $5/2$ kernel) used in materials science[8] and the BO literature[11] since it is good at fitting functions with steep gradients. We iterated the routine after the initial MBE growth with five random initial growth parameters and RRR measurements. First, GPR was updated using the data set at the time (Fig. 5d). Subsequently, to assign the value of the growth parameter in the next run, we calculated the EI[50].

**Author Contributions**
Y.K.W. and T.O. conceived the idea, designed the experiments, and directed and supervised the project. T.O. and Y.K.W. implemented the Bayesian optimization algorithm. T.O. performed simulations with the virtual data. Y.K.W. carried out the ML-MBE growth and characterization of the samples. Y.K.W. and T.O. co-wrote the paper with input from all authors.

**DATA AVAILABILITY**
The data that support the findings of this study are available from the corresponding authors upon reasonable request.

**Figures and figure captions**

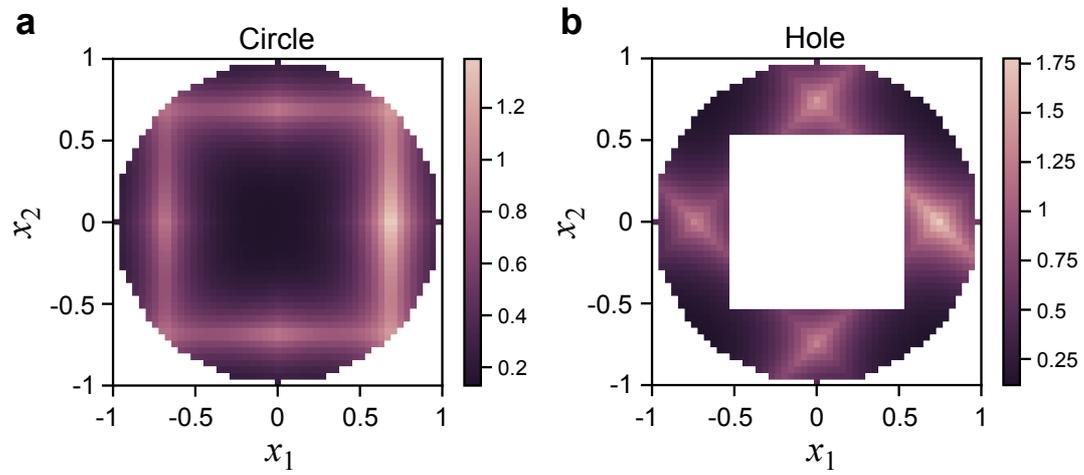

**Fig. 1. Visualization of functions used in the simulated experiments.** Search spaces are two-dimensional [-1, 1] boxes for both the Circle (**a**) and Hole (**b**) functions. White areas indicate experimental failure regions.

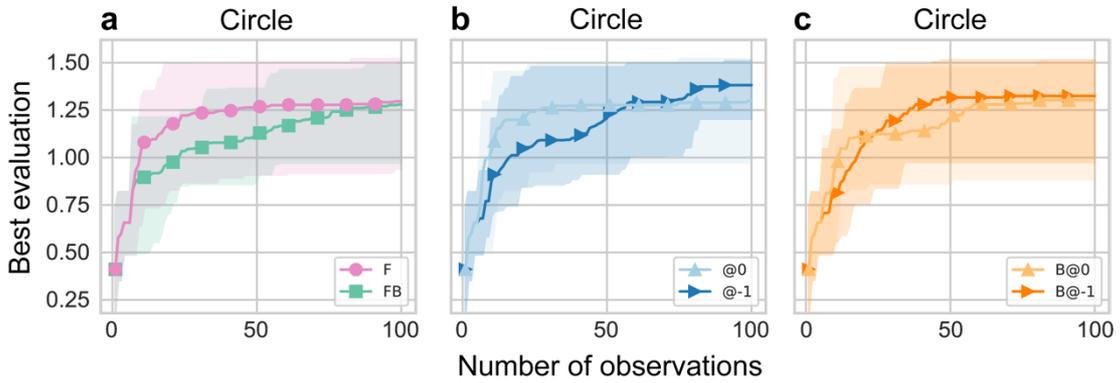

**Fig. 2. Parameter search results for the Circle function.** Best evaluation values averaged over five runs as a function of the number of observations. Shaded areas indicate the range of the best and worst values of five runs. **a** Methods F and FB that adaptively replace failed evaluations by the floor padding. **b** Baseline methods that use a constant value for the failed evaluations. Method @0 used the fixed value of 0 when the experiment failed, whereas @−1 used −1. **c** Methods with classifier while padding a constant to failures. B@0 used a value of 0 as the evaluation when the experiment failed, whereas B@−1 used −1.

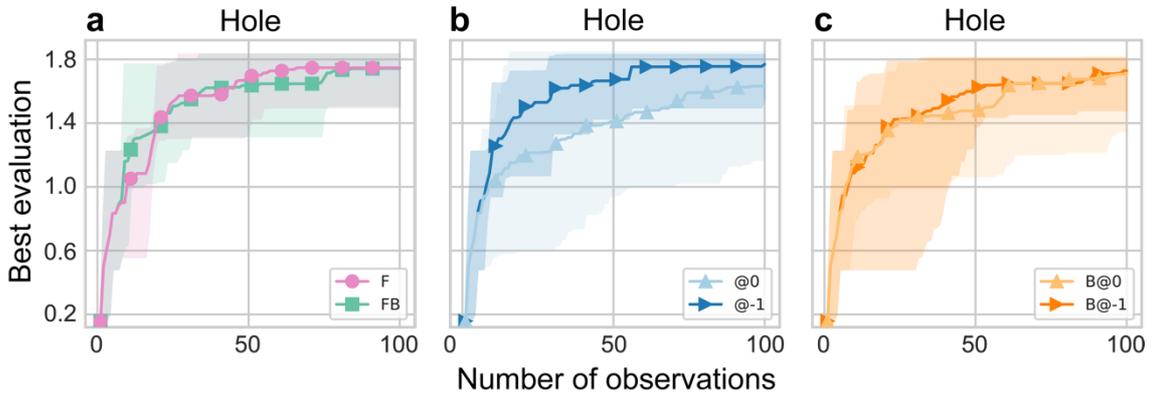

**Fig. 3. Parameter search results for the Hole function.** Best evaluation values averaged over five runs as a function of the number of observations. Shaded areas indicate the range of the best and worst values of five runs. **a** Methods F and FB that adaptively replace failed evaluations by the floor padding. **b** Baseline methods that use a constant value for the failed evaluations. Method @0 used the fixed value of 0 when the experiment failed, whereas @−1 used −1. **c** Methods with classifier while padding a constant to failures. B@0 used a value of 0 as the evaluation when the experiment failed, whereas B@−1 used −1.

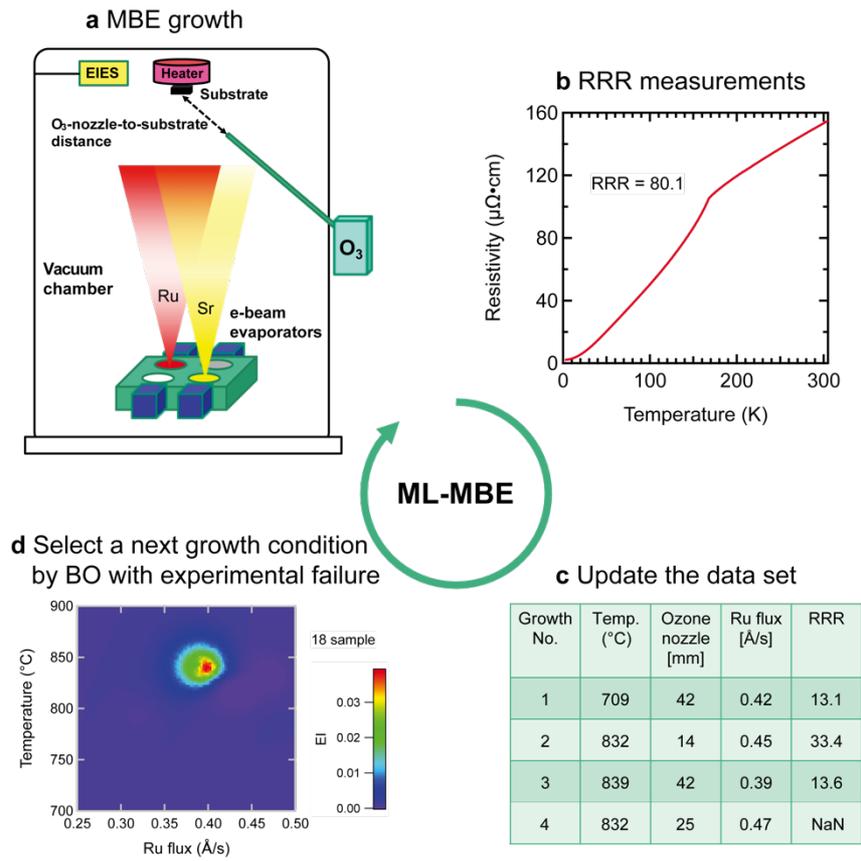

**Fig. 4. Flow of ML-MBE growth using BO with experimental failure. a** Schematic illustration of our multisource oxide MBE setup. EIES: Electron Impact Emission Spectroscopy. **b** Resistivity vs. the temperature curve of the $SrRuO_3/DyScO_3$ film with a RRR of 80.1, as an example. **c** Growth conditions for four samples, as an example. **d** Two-dimensional plots of EI values at the $O_3$-nozzle-to-substrate distance of 22.5 mm obtained from the collected data for 18 samples, as an example.

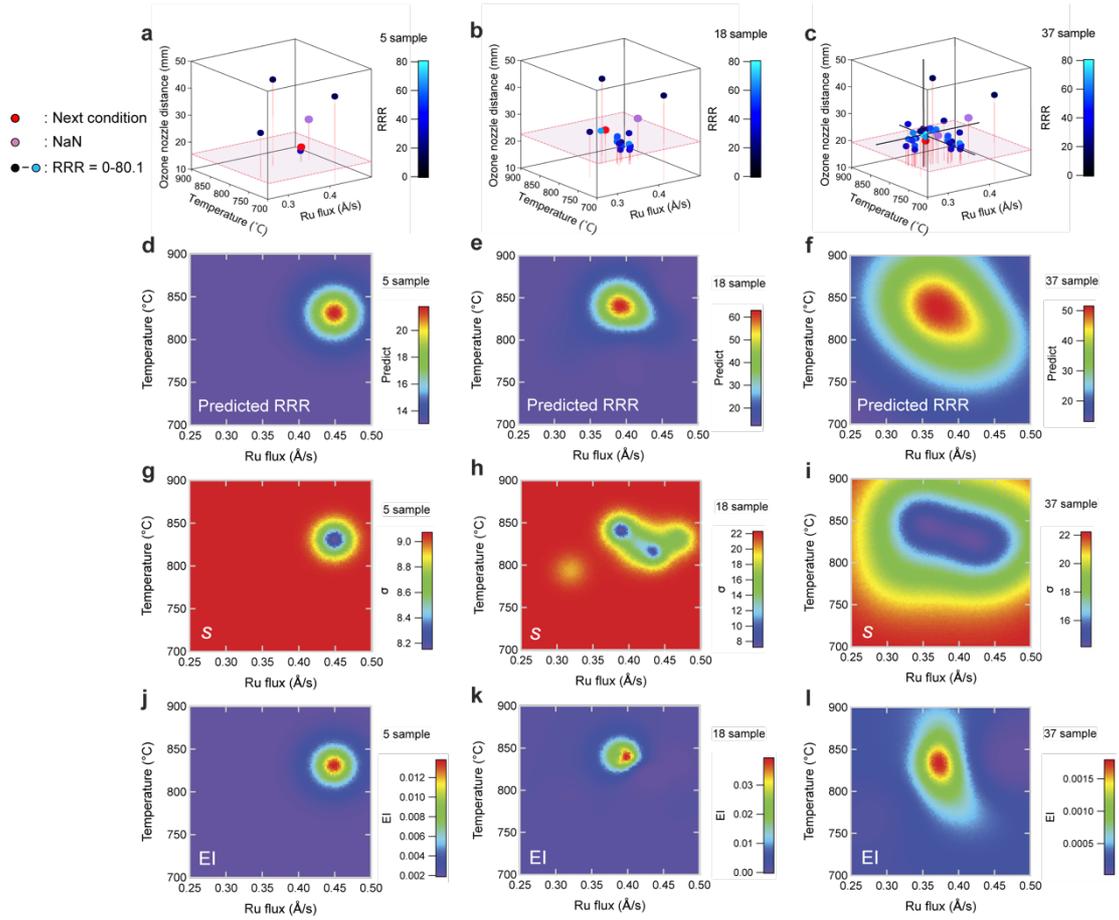

**Fig. 5. Experimental RRR values and prediction results from GPR. a-c,** Experimental RRR values in the three-dimensional growth parameter space for 5 (**a**), 18 (**b**), and 37 (**c**) samples. The purple spheres indicate the NaN points at which the SrRuO$_3$ phase was not obtained. The red spheres indicate the next conditions at which the highest EI values were obtained. The red planes indicate the cutting plane of the O$_3$-nozzle-to-substrate distance, at which the highest EI value was obtained. **d-l,** Two-dimensional plots of predicted RRR values (**d-f**), $\sigma$ values (**g-i**), and EI values (**j-l**) at the O$_3$-nozzle-to-substrate distance of 15.5 mm (**d**,**g**,**j**), 22.5 mm (**e**,**h**,**k**), and 19. 5 mm (**f**,**i**,**l**), which were obtained from the collected data for 5 (**d**,**g**,**j**), 18 (**e**,**h**,**k**), and 37 (**f**,**i**,**l**) samples, respectively. The O$_3$-nozzle-to-substrate distance was that at which the highest EI value was obtained.

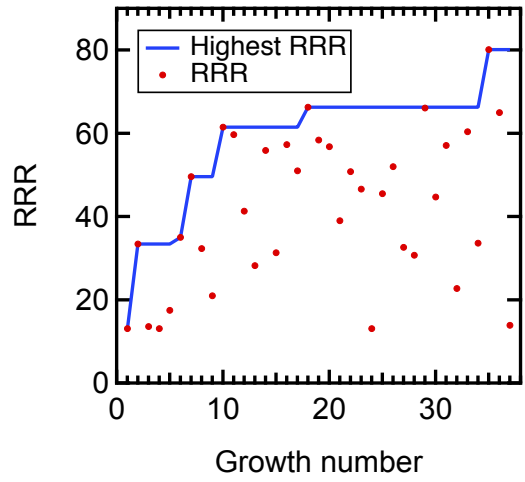

**Fig. 6. ML-MBE Optimization results.** Highest experimental RRR and RRR values plotted as a function of the growth number.

|          | Floor padding | Binary classifier |
|----------|:-------------:|:-----------------:|
| Baseline |               |                   |
| F        | ✓             |                   |
| B        |               | ✓                 |
| FB       | ✓             | ✓                 |

**Table 1** Compared methods. The floor padding trick replaces missing evaluation with the worst value adaptively. The binary classifier takes into account the probability of experimental failure.

# Supplementary Information

# Bayesian optimization with experimental failure for high-throughput materials growth

Yuki K. Wakabayashi,[1,*,†] Takuma Otsuka,[2,*,‡] Yoshiharu Krockenberger,[1] Hiroshi Sawada,[2] Yoshitaka Taniyasu,[1] and Hideki Yamamoto[1]

[1]*NTT Basic Research Laboratories, NTT Corporation, Atsugi, Kanagawa 243-0198, Japan*
[2]*NTT Communication Science Laboratories, NTT Corporation, Soraku-gun, Kyoto 619-0237, Japan*

[*]These authors contributed equally to this work.
[†]Corresponding author: yuuki.wakabayashi.we@hco.ntt.co.jp
[‡]Corresponding author: takuma.otsuka.uf@hco.ntt.co.jp

---
**Algorithm** Bayesian optimization with experimental failure
---
**Input:** initial data $\mathcal{D} := \{(\mathbf{x}_i, y_i)\}_{i=1,\ldots,\text{\# of init}}$, budget $N$
**Output:** best parameter $\mathbf{x}_i$ such that $i = \text{argmax}_{1 \leq i \leq N}\, \tilde{y}_i$
    **while** $|\mathcal{D}| < N$ **do**
        $n \leftarrow |\mathcal{D}| + 1$
        preprocess data into $\widetilde{\mathcal{D}} = \{(\tilde{\mathbf{x}}_n, \tilde{y}_n)\}_{i=1}^{n-1}$ from $\mathcal{D}$ by Eqs. (1) and (2)
        fit Gaussian process using $\widetilde{\mathcal{D}}$ to calculate $m$ and $s^2$ in Eqs. (3) and (4)
        find next parameter $\tilde{\mathbf{x}}_n = \text{argmax}_{\tilde{\mathbf{x}}'}\, a_{\text{EI}}(\tilde{\mathbf{x}}')$
        rescale to original parameter $\mathbf{x}_n$ from $\tilde{\mathbf{x}}_n$
        obtain evaluation for parameter $\mathbf{x}_n$ as $y_n$
        add to raw data $\mathcal{D} \leftarrow \mathcal{D} \cup \{(\mathbf{x}_n, y_n)\}$
    **end while**
---

**Supplementary Fig. 1 Pseudocode of Bayesian optimization.**

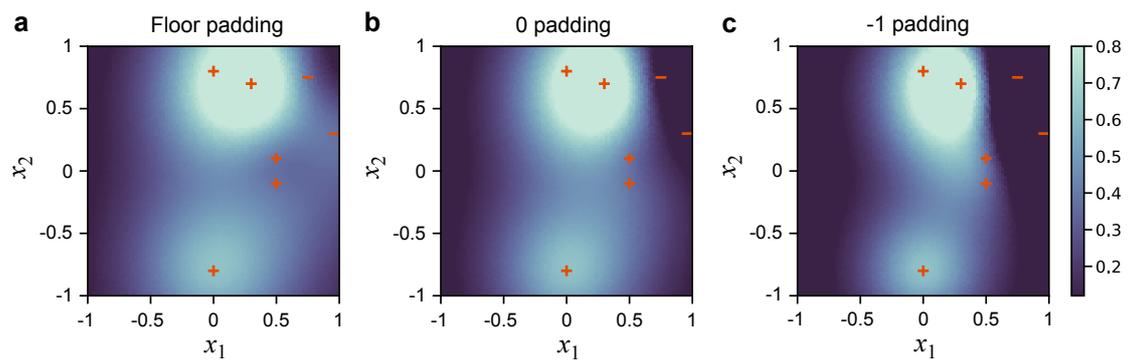

**Supplementary Fig. 2 Effect of padding with various values.** The predicted mean of the Gaussian process for the Circle function after seven observations with five successful '+' and two failed '−' observations. The evaluations of '−' marks are replaced by the worst evaluation of '+' marks (**a**), 0 (**b**), and −1 (**c**), respectively.

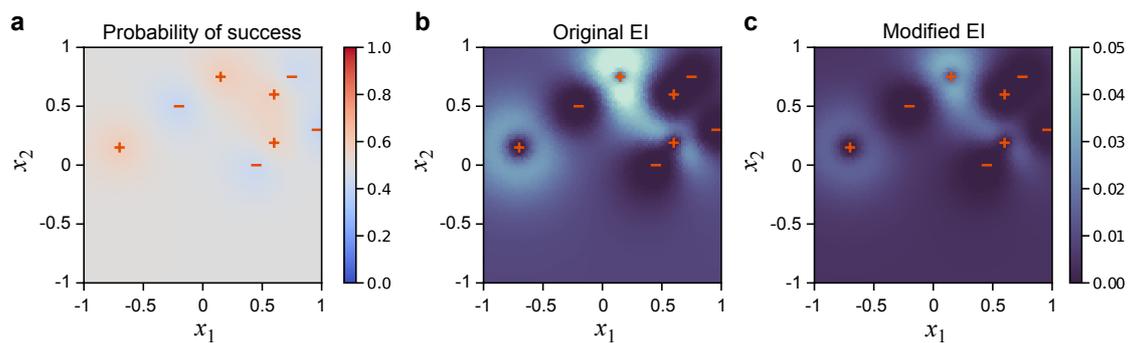

**Supplementary Fig. 3 Effect of binary classifier on expected improvement.** The prediction by method FB from eight observations of the Hole function with four successful '+' and four failed '−' observations. (**a**) Predicted success probability. (**b**) The original EI function used by method F. (**c**) Modified EI function of method FB, given as the product of (**a**) and (**b**).